\documentclass{emulateapj}
\usepackage{natbib}
\usepackage{graphicx}

\usepackage{epsfig,natbib}

\newcommand{\be}{\begin{equation}}
\newcommand{\ee}{\end{equation}}
\newcommand{\ba}{\begin{eqnarray}}
\newcommand{\ea}{\end{eqnarray}}
\def\lesssim{\mathrel{\hbox{\rlap{\hbox{\lower4pt\hbox{$\sim$}}}\hbox{$<$}}}}
\def\grtsim{\mathrel{\hbox{\rlap{\hbox{\lower4pt\hbox{$\sim$}}}\hbox{$>$}}}}

\begin{document}

\title{The galaxy-galaxy lensing contribution to the cosmic shear two point function}

\author{Sarah Bridle and Filipe B. Abdalla}
\affil{Department of Physics and Astronomy, University College London,
Gower Street, London, WC1E 6BT, UK}

\begin{abstract}
We note that galaxy-galaxy lensing by non-spherical galaxy halos
produces a net anti-correlation between the shear of background
galaxies and the ellipticity of foreground galaxies.
This anti-correlation would contaminate the tomographic cosmological
weak lensing two point function if the effect were not taken into account.
We compare the size of the galaxy-galaxy lensing contribution to
the change in the cosmic shear two-point cross-correlation function due
to a change in the dark energy equation of state $w$ of 1\%. We find them
comparable on scales $\lesssim 5'$ for NFW galaxy profiles, and out
to much larger scales for SIE profiles.
However the galaxy-galaxy lensing signal has a characteristic
spatial and redshift pattern which should allow it to be removed.
\end{abstract}



\section{Introduction}
\label{sec:introduction}

Cosmic shear shows great promise for testing the cosmological
model and measuring cosmological parameters.
It measures the gravitational bending (lensing) of light by all the intervening
mass in the Universe. In cosmic shear, distant (background) galaxies
provide a convenient
screen of objects with potentially simple statistical properties,
from which the light bending can be inferred.
The distortion depends on the lens geometry and thus on the curvature
and expansion history of the Universe; it also depends on the distribution
of matter, which itself depends on most aspects of the cosmological model.

Cosmic shear was first detected in 2000
\citep{wittmanea00,baconre00,vanwaerbekeea00,kaiserwl00}
and has been used to
constrain cosmological
parameters in subsequent surveys \citep[recently][]{hoekstraea05}.
Cosmic shear has the potential to become the most powerful probe of
cosmology because it observes a non-Gaussian
three dimensional field in the local
Universe, where the mysterious dark energy dominates.
However
a number of details inevitably remain to be
investigated further whilst planning for future experiments.

The simplest statistic is the two point angular correlation function
\citep[or integrals of this quantity such as the power spectrum or
aperture mass,][]{schneider96}.
If the background galaxies can be separated into populations at
different redshifts the power of this statistic is
greatly increased
by cross-correlating galaxies in different redshift bins
\citep{hu99}.
This technique will dominate for future surveys which aim to
measure the equation of state of the dark energy, or determine
a new gravitational theory.

Cosmic shear is particularly simple if distant background galaxies
can be assumed to have random orientations. The distortion
by gravitational lensing can then be extracted statistically, for
example by averaging over galaxy orientations in a given patch of sky.
Unfortunately it is unlikely to be straightforward since when galaxies
form they tend to align pointing towards dark matter concentrations
due to tidal interactions.
This leads to two complicating effects:
(i) neighbouring galaxies at a given redshift are intrinsically aligned
\citep{hoyle49,peebles69,heavensp88,crittendennpt01}
and
(ii) a pair of galaxies at two different redshifts have correlated observed
ellipticities because a dark matter concentration close to the nearer
object may tidally align the nearer object \emph{and}
simultaneously gravitationally
lens the more distant object \citep{hiratas04}.

These intrinsic alignment effects are complicated because they
require an understanding of tidal alignments of galaxies, however
they do have a distinctive signature and can therefore be removed
without requiring a detailed understanding of galaxy formation
\citep{kings02,kings03,heymansh03,king05}.

In this paper we present another complicating effect
which is much simpler than the intrinsic alignment
effects:
cross-correlation of a pair of galaxies at different
redshifts produces a contribution to the usual cosmic
shear two-point statistic even if there are no tidal effects.
This is because the nearer galaxy will gravitationally
lens the more distant galaxy (galaxy-galaxy lensing).
This effect produces a net anti-correlation
in the realistic case when the nearer galaxy (i) is
not circularly symmetric
and (ii) has a correlation between the asymmetry of the light
and the mass.
In this paper we present this new factor and
quantify its effect for elliptical dark matter halos
for which mass and light have the same ellipticity and orientation.

First we review ellipticity and shear correlation function notation.
We then illustrate the galaxy-galaxy lensing contribution to the
ellipticity correlation function and calculate the
effect due to a population of galaxy lenses at a fixed redshift.
We investigate how this effect depends on the source and lens
redshift.
Finally we compare this to the size of the tomographic cosmic shear cross-correlation
signal.

\section{The ellipticity correlation function}
\label{sec:ellcorfun}

\begin{figure}
\epsfig{file=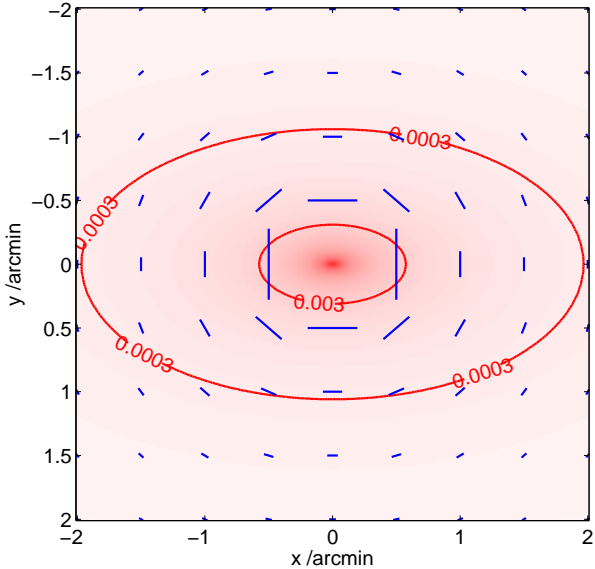,width=8cm,angle=0}
  \caption{
  Shading and contours show the convergence for an
  elliptical NFW mass distribution ($e=0.3$, $M_{200}=1.2\times 10^{12}
  M_{\odot}/h$)
  at a redshift of 0.3 with sources at redshift 0.8.
  Lines show the resulting shear map.
  Note that sticks on the x axis are larger than sticks on the y
  axis for same distance from the lens center.
  }
  \label{fig:map}
\end{figure}

The two-point correlation function of the shear field $\gamma({\bf x})$
at positions separated by an
angle $\theta$ on the sky is given by
$\xi_{\gamma}=\langle \gamma \gamma^*\rangle$,
averaging over all pairs of angular separation $\theta$
(We use complex notation for shears $\gamma = \gamma_1 + i \gamma_2$
and similarly for ellipticities.)
For a concise introduction to cosmic shear see
\cite{refregier03}; see also \cite{mellier99} and \cite{bartelmanns01}.

Cosmic shear aims to measure this correlation function by
measuring observed ellipticities of distant galaxies
$e^o$ and taking into account how intrinsic (pre-shear) ellipticities
$e^i$ are modified by shear. For small shears this reduces to
$e^o = e^i + \gamma$; we define $e \equiv (a-b)/(a+b)$ throughout.
An estimate of the shear two point correlation function is obtained from
the observed ellipticity correlation function
$\xi_{e}=\langle e^o_{\rm s} e^{o*}_{\rm d}\rangle$
where $e^o_{\rm s}$ is the observed ellipticity of a distant galaxy
${\rm s}$ and $e^o_{\rm d}$ is that of a nearer galaxy ${\rm d}$.
Therefore we can expand the correlation
functions to find $\xi_{e} = \xi_{\gamma} + \xi_{\gamma e}$,
where the shear-ellipticity correlation is given by
 $\xi_{\gamma e} =\langle \gamma_{\rm s} e^{i*}_{\rm d} \rangle$
if we ignore intrinsic alignments
($\langle e^{i}_{\rm s} e^{i*}_{\rm d} \rangle=0$)
and use $\langle e^{i}_{\rm s} \gamma_{\rm d} \rangle=0$.

\cite{heymanswhvv06} calculated the shear-ellipticity correlation function
numerically using n-body simulations. The aim of that work was
to quantify the intrinsic alignment - shear correlation presented
by \cite{hiratas04}.
Their results should in fact contain a mixture of the
intrinsic alignment - shear correlation and the galaxy-galaxy lensing
signal we present here. However, they do not discuss the
distinction between the two effects,
or specifically state that the galaxy-galaxy lensing contribution
might be significant.
Here we present simple analytical and numerically integrated results
to quantify only the galaxy-galaxy lensing contribution.

We assume a concordance $\Lambda$CDM cosmology throughout, with
parameters taken from \cite{spergelea06}:
Hubble constant $H_0=73$ km s$^{-1}$ Mpc$^{-1}$, $\Omega_{\rm m}=0.238$,
$\Omega_{\rm DE}=1-\Omega_{\rm m}$, $\Omega_{\rm b}=0.047$,
$\sigma_8=0.74$, dark energy equation of state $w=-1$ except
where otherwise stated.
We assume a flat universe with a scale invariant primordial power spectrum.

\begin{figure}
\epsfig{file=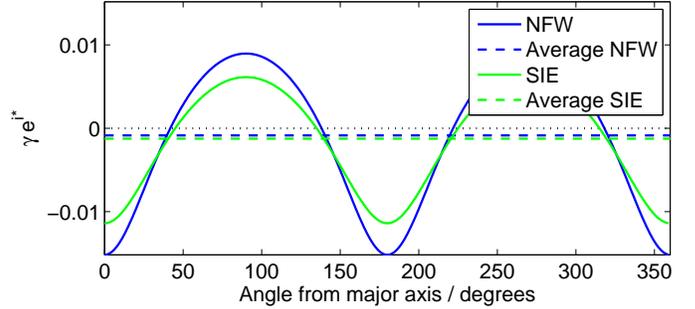,width=9cm,angle=0}
  \caption{
  \emph{Solid lines:} the shear-ellipticity correlation ($\gamma e^{i*}$)
  at an angular separation of $0.1'$ for a single lens.
  \emph{Larger amplitude solid line:} For the elliptical NFW shown in Fig 1.
  \emph{Smaller amplitude solid line:} For an SIE with the same $M_{200}$.
  \emph{Upper dashed line:} Outer solid line averaged over angle.
  \emph{Lower dashed line:} Inner solid line averaged over angle.
  In both cases the variation mostly cancels leaving a negative
   shear-ellipticity correlation.
  Note that averaging over many background galaxies with many position
  angles
  is equivalent to integrating under the sine-like curve and the
  effect persists.
In this case the shear-ellipticity correlation is simply a rescaling
of the shear by a factor of $e = 0.3$.
  }
  \label{fig:map_angle}
\end{figure}

\section{The galaxy-galaxy lensing contribution}
\label{sec:methodology}

In this Section we calculate the galaxy-galaxy lensing contribution
to $\xi_e$, given by $\xi_{\gamma e}$, as discussed above.
To estimate this quantity we first calculate the  signal
for a single elliptical lens averaged over background galaxies
at a fixed angular distance, and then average over a population of lenses.
We assume that the lens light has the same ellipticity and orientation
as the lens mass.

By default we consider an NFW \citep{nfw} mass profile,
calculating the projected mass from the equations given in \cite{wrightb00}
and \cite{bartelmann96}.
We use $M_{200}$,
the mass enclosed within the radius at which the
density is 200 times the mean density of the Universe, for consistency
with simulations.
We derive the concentration parameter, $c$, as a function of
$M_{200}$ using Eq. 12 of \cite{seljak00}
with $\beta=-0.15$, as appropriate for an NFW model.

We calculate the shear for an elliptical mass distribution using
the equations in \cite{keeton01}
and \cite{schramm90}.
Note that this is not the same as calculations using elliptical
potentials, which give dumbbell shaped mass distributions
 \citep[e.g.,][]{kassiolak93}.
The projected mass distribution is squashed and stretched to have
elliptical isodensity contours a factor $f$ smaller (larger) along
the minor (major) axes,
as compared to the corresponding spherical mass distribution.

The shear map for an elliptical NFW lens of ellipticity
$e =0.3$ aligned along the $x$ axis
is shown in Fig.~\ref{fig:map}.
The shading and contours show the convergence map
(projected mass density in units of the critical lens
density).
The overlaid shear sticks show two particularly interesting
features:
(i) the shear on the major axis of the lens is larger than
that on the minor axis, for a given angular separation from the lens
center;
(ii) the shear 45 degrees around from the major axis is
approximately tangential to the center of the lens.
These two features do depend on the details of the mass profile,
but are general for relevant radii for an elliptical NFW profile, and also
for a singular isothermal ellipsoid (SIE)
 \citep[for which the shear is always exactly tangential and its
 amplitude follows the mass, see][]{kassiolak93,kormannsb94}.

These two characteristics point towards our main result: that there
is a net anti-correlation between lens ellipticity and the resulting
shear of background galaxies. This arises because
(i) the shears on the major and minor axes cancel out, but only
partially, leaving a shear which is perpendicular to the major axis
of the lens;
(ii) the shear sticks in-between the major and minor axes do not
remove this anti-correlation (for example this could have happened had the
shear sticks been aligned along iso-density contours).

This is quantified in detail in Fig.~\ref{fig:map_angle}, which
shows the shear-ellipticity correlation for a lens at $z=0.3$ with a
background source at $z=0.8$, but as a function of $\theta$
around the lens center (consider moving around a concentric
circular annulus
in Fig.~\ref{fig:map}). The integral under this curve shows the net effect
of averaging over galaxies.

The shape of the curve is non-trivial, but on average is below
zero, as shown by the dashed lines.
The larger amplitude oscillation is for the elliptical NFW shown in Fig.~\ref{fig:map},
and the smaller oscillation is for an SIE with the same $M_{200}$.
They both show the same qualitative
effect at the radius used for the figure (0.1 arcmin).
At larger radii, for the SIE, the average shear-ellipticity correlation
remains a similar fraction of the maximum due to the scale independence
of the SIE. For the NFW profile, the average becomes a smaller fraction
of the maximum due to the steepening of the NFW profile with radius.

\begin{figure}
\epsfig{file=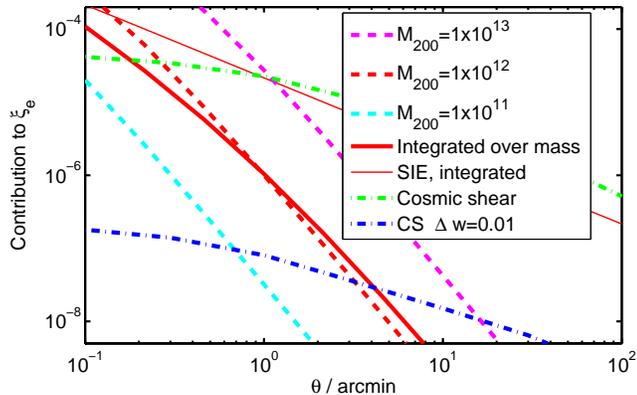,width=8.5cm,angle=0}
  \caption{
\emph{Dashed lines:}
The galaxy-galaxy lensing contribution to the ellipticity correlation
function for lenses of varying mass (from lowest to highest:
$1\times10^{11} M_{\odot}/h$, $1\times10^{12} M_{\odot}/h$
and $1\times10^{13} M_{\odot}/h$; $|e|=0.2$, $z_{\rm d}=0.3$, $z_{\rm s}=0.8$).
(Absolute values are shown; The correlation is negative at all scales and masses.)
\emph{Thick solid line:}
Averaged over an $R<24$ population with rms ellipticity of each component
0.16 ($z_{\rm d}=0.3$, $z_{\rm s}=0.8$).
\emph{Thin solid line:}
As for thick solid line but for a SIE profiles instead of elliptical NFWs.
\emph{Upper dash-dot line:}
The cosmic shear two point correlation function between
redshift 0.3 and 0.8.
\emph{Lower dash-dot line:}
The difference between cosmic shear correlation functions with dark energy
equation of state values $w=-1$ and $w=-0.99$.
}
  \label{fig:cf cosmic shear}
\end{figure}

The middle dashed line in Fig.~\ref{fig:cf cosmic shear} shows
this
average quantity
as a function of angular separation
for a single lens of mass $10^{12} M_{\odot} /h$
and ellipticity $0.2$.
The other dashed lines show how the effect depends on lens mass.

So far we have shown that for a single foreground lens there is an
anti-correlation between shear and lens ellipticity. We now compare
the cosmic shear tomographic cross correlation ($\xi_{\gamma}$)
with the average of $\xi_{\gamma e}$ over all foreground lenses.
Qualitatively: the shear-ellipticity correlation for a single lens is
a scalar quantity, and therefore is independent of the angle of the
coordinate system. As a result, on averaging over many isolated
lenses at different angles to each other, the anti-correlation
will be preserved: we average many negative numbers
($\gamma e^{i*}$ for each lens) together and still
obtain a negative number.
This assumes that the foreground lenses are isolated from
each other, and that there is no additional lensing from
structures the foreground lenses are aligned with (i.e.
ignores the intrinsic alignment -- shear correlation).

We now quantify the effect of averaging over many isolated
lenses at a given redshift.
First we assume all lenses have the same mass and consider
just the variety of ellipticities.
We average over a population of lens ellipticities,
drawing each ellipticity component from a Gaussian of width
0.16, and calculating the absolute ellipticity.
We find that the shear-ellipticity correlation scales roughly as
the square of the lens ellipticity. This implies that the
mean shear scales with lens ellipticity.
The shear-ellipticity correlation function, after averaging
over this population, is roughly equal to that for a single
lens of ellipticity $0.23$. As expected from the squared
dependence, this is slightly higher than the mean absolute ellipticity
of the lenses ($0.20$).

We also average over the lens population mass distribution expected for a
survey with shear measurements complete to an apparent magnitude
limit of 24 in the SDSS r filter.
We obtain number densities from \cite{shetht99}
calculated at the lens redshift, down to a limiting mass $M_{\rm lim}$
which corresponds to the magnitude limit.
This limiting mass was determined by
(i) calculating from the mass function the number of objects per unit
volume integrated down to the limiting mass
(ii) comparing this with number density of observed galaxies derived
from COMBO-17 luminosity functions \citep{wolfea03}, using the
numbers in Table~1 of \cite{blakeb04}.
We find $M_{\rm lim}=3\times10^{10} M_{\odot} /h$, which gives a
mean mass of $4\times10^{11} M_{\odot} /h$.

We investigated the effect of averaging over mass by considering
the contribution to $\xi_e$ for a fixed lens ellipticity.
At 4 arcmin the $\xi_e$ contribution averaged over mass
is equal to that for a single lens of mass
$1.2\times10^{12} M_{\odot} /h$.
This is bigger than the mean mass of the objects because the
dependence of the shear-ellipticity correlation is roughly
$M^{1.5}$ at this angular scale.

The thick solid line in Fig.~\ref{fig:cf cosmic shear} shows the
contribution to the ellipticity correlation function after
averaging over mass and ellipticity.
It is slightly more shallow than the lines at constant mass
since larger masses dominate at larger angular separations
due to their smaller concentration parameters.
This is slightly amplified by the fact that larger ellipticities
contribute more at larger radii, and larger ellipticities have
a larger shear-ellipticity correlation.

To estimate the dependence on lens profile we repeat the above
calculations using a population of SIE lenses, with velocity
dispersions calculated analytically from $M_{200}$.
The resulting contribution is shown by the thin solid line in
Fig.~\ref{fig:cf cosmic shear}, and is much higher than the NFW
result, likely mainly because of the shallower density profile
at large radii ($\propto r^{-2}$ instead of $\propto r^{-3}$).

All of the above calculations assume a lens redshift of
$z_d=0.3$ and a background redshift of $z_s=0.8$.
The solid lines in Fig.~\ref{fig:z dep} show how the signal at 4 arcmin
depends on lens and source redshift.
In general the signal is larger at lower lens redshifts and decreases
rapidly to zero as the lens approaches the source.
We expect the signal to be smaller as the lens approaches the source
because of the geometry of gravitational lensing.
The fact that it drops so quickly, and in fact for $z_s=1.2$ rises
slightly first, is because at higher redshifts a magnitude limited
survey contains more massive objects, that are more effective at
lensing.
At low redshift, galaxies have a larger angular size so for a
given angular separation we start to probe into the center of the
NFW profile where the profile slope is more constant with radius.
However, on averaging over redshift the lower redshift contribution would
become smaller since there will be fewer objects in the light cone.
Also there is little cosmic shear signal at low redshift so these
galaxies could be ignored.

Finally we investigated the effect of survey depth, and find
that for a given angular scale, fixing $z_d=0.3$ and $z_s=0.8$,
the signal was around a factor 2.5 larger if $r<22$, and
a factor approximately 4 smaller if $r<27$, depending on the
extrapolation of the luminosity function.
This is to be expected because deeper surveys contain many
less massive objects.

\section{Discussion}

\begin{figure}
\epsfig{file=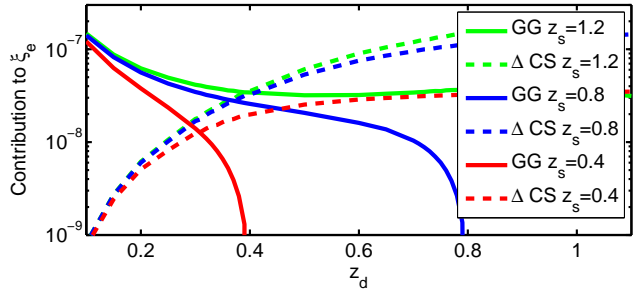,width=8.5cm,angle=0}
  \caption{
\emph{Solid lines:}
Dependence of the galaxy-galaxy lensing shear-ellipticity
correlation at 4 arcmin on lens redshift for source redshifts 0.4, 0.8 and 1.2
(lower, middle and upper lines respectively).
\emph{Dashed lines:} Cosmic shear cross-correlation between a
sample at a redshift $z_l$ and samples at redshifts $z_s= 0.4$,
0.8 and 1.2 (lower, middle and upper lines respectively);
for a change in $w$ of 0.01
(the result for $w=-0.99$ is subtracted from the result for $w=-1$).
  }
  \label{fig:z dep}
\end{figure}

In Figs.~\ref{fig:cf cosmic shear} and~\ref{fig:z dep}
we compare the galaxy-galaxy lensing contribution to the
ellipticity correlation function with that from cosmic shear
cross-correlation tomography.
We calculate the non-linear matter power spectrum as a function of
redshift using CAMB~\citep{lewiscl00} with the HaloFit
\citep{smithea03} option. This is converted into the cosmic shear
cross power spectrum using equations from \cite{hu99}.
We assume delta function redshift distributions corresponding
to the galaxy-galaxy lens and source redshifts.

The cosmic shear correlation function is flatter than the
galaxy-galaxy lensing contribution
Above $\sim0.2'$ the galaxy-galaxy lensing
contribution is smaller than the cosmic shear signal for NFW lenses.
However it must be much smaller than the cosmic shear signal
if it is not to interfere with cosmological parameter analyses.

To illustrate this we also show the change in cosmic shear signal
for a small change ($\Delta w=0.01$) in the equation of state of
dark energy, $w$,
leaving the other cosmological parameters (including $\sigma_8$)
fixed.
The galaxy-galaxy lensing contribution is not negligible
below 4 arcmin for NFW galaxy profiles. We find that this scale will
strongly depend on the lens profile since it is greater than
30 arcmin for SIE profiles.

We assumed that the lens ellipticity is the same size and orientation
as the mass ellipticity. This is consistent with measurement attempts
\citep{hoekstrayg04,mandelbaumea06}, but the signal to noise is low.
If the mass ellipticity were a factor $f$ smaller than the light
then this would simply scale the galaxy-galaxy lensing contribution
by $f$. If there is a misalignment then the effective $f$ on stacking
would be a first approximation to the change.
\cite{heymanswhvv06} consider a more sophisticated alignment
model which can reduce the signal by a factor of roughly three.

It would be interesting to see whether the contribution is increased or
reduced on considering clustering of the lens galaxies.
Further, it is not clear whether deviations from ellipsoidal symmetry
of the lens (such as substructures) would average out.

A rough comparison of Fig.~\ref{fig:cf cosmic shear} with
\cite{heymanswhvv06} (Table 2)
indicates that at 1 arcminute the galaxy-galaxy NFW lensing
contribution is a significant fraction of the total shear-intrinsic
alignment correlation, whereas at 10 arcmin the shear-intrinsic
alignment correlation is dominated by tidal alignments. Note that
this comparison is only approximate since \cite{heymanswhvv06} integrate
over lens redshift.

Fortunately it may be relatively easy to remove the galaxy-galaxy
lensing contribution using the characteristic tangential nature of
galaxy-galaxy lensing. For example the ellipticity two-point
function could be measured as a function of two dimensional
separation $(\theta_1, \theta_2)$, where the position angle $\alpha$
is measured from the major axis of
the foreground lens. \citep[This ``stack and rotate" method was
advocated by][to measure ellipticity of galaxy dark matter
halos.]{natarjanr00} The galaxy ellipticity and profile could then
be fitted simultaneously with extracting the cosmic shear two point
function. Alternatively one could use the method of \cite{king05},
which uses the redshift signature of the effect.
Also \cite{heymanswhvv06} suggest removing the most luminous galaxies
in the lower redshift slice.

This effect should also ultimately be taken into account for higher
order statistics of cosmic shear, since they probe smaller scales
than the two-point statistic \citep[see ][for a calculation for the
three point function]{how01}.
Furthermore the galaxy-galaxy flexion signal could be a bigger
contaminant of cosmic flexion because the latter is more
sensitive than cosmic shear at small angular scales.

\acknowledgements

We thank
Keith Biner,
Masahiro Takada,
Peter Schneider,
Virginia Corless,
Eduardo Cypriano,
David Sutton,
Phil Marshall,
Richard Massey,
Fergus Simpson,
Adi Nusser
and
Lindsay King
for helpful discussions.
SLB acknowledges support from a Royal Society University Research Fellowship.

\end{document}